\documentclass[aps,prl,twocolumn,showpacs,superscriptaddress]{revtex4-1}

\usepackage{amssymb,amsmath,graphicx,graphics}

\begin{document}

\title{Tunable frequency-up/down conversion in gas-filled hollow-core photonic crystal fibers}
\author{Mohammed F. Saleh}
\affiliation{School of Engineering and Physical Sciences, Heriot-Watt University, EH14 4AS Edinburgh, UK}
\affiliation{Department of Mathematics and Physics Engineering, Alexandria University, Alexandria, Egypt}
\author{Fabio Biancalana}
\affiliation{School of Engineering and Physical Sciences, Heriot-Watt University, EH14 4AS Edinburgh, UK}

\begin{abstract}
Based on the interplay between photoionization and Raman effect in gas-filled photonic crystal fibers, we propose a new optical device to control frequency-conversion of ultrashort pulses. By tuning the input-pulse energy, the output spectrum can be either down-converted, up-converted, or even frequency-shift compensated. For low input energies, Raman effect is dominant and leads to a redshift that increases linearly during propagation. For larger pulse energies, photoionization starts to take over the frequency conversion process, and induces a strong blueshift. We have found also that the fiber-output pressure can provide an additional degree of freedom to control the spectrum shift.
\end{abstract}

\maketitle

Hollow-core (HC) photonic crystal fibers (PCFs) have been developed in the last decade to explore nonlinear light-matter interactions, with a flexibility far-beyond the all-solid optical fibers \cite{Russell03,Russell06}. HC-PCFs with Kagome-style lattice have offered unbeatable advantages to host strong nonlinear interactions between light and different gaseous media \cite{Russell14}. These microstructures are characterized by having wide transparency range and a pressure-tunable zero dispersion wavelength (ZDW) in the visible range \cite{Travers11}. The abundance of gases with different properties have led to the observation of various novel nonlinear phenomena \cite{Benabid02a,Heckl09,Joly11,Hoelzer11b,Saleh11a,Saleh12,Chang13}. For instance, efficient deep-ultraviolet radiation has been observed in argon-filled HC-PCFs  \cite{Joly11}, while ultrabroadband supercontinuum generation spanning from the vacuum-UV to  the mid infrared regime is obtained using hydrogen-filled HC-PCFs \cite{Belli15}.

Raman self-frequency redshift, which continuously downshifts the central frequency of a single pulse during propagation, has been first observed in solid-core fibers  \cite{Dianov85,Mitschke86,Bulushev91}, and later also in molecular gases \cite{Korn98}. Photoionization-induced self-frequency blueshift due to plasma generation have been predicted and demonstrated in HC-PCFs filled by argon gas  \cite{Hoelzer11b,Saleh11a,Chang13}. The core of this study  is to investigate the non-trivial interplay between the Raman and photoionization nonlinear effects in HC-PCFs filled by Raman-active gases, and demonstrate its potential in designing novel photonic devices.

We consider the propagation of an ultrashort pulse in a HC-PCF filled by a Raman-active gas. The  evolution of the pulse electric field can be accurately described via solving the unidirectional pulse propagation equation (UPPE) that does not require the slowly envelope approximation \cite{Kolesik04,Kinsler10},
\begin{equation}
i\dfrac{\partial\tilde{E}}{\partial z}=i\left( \beta\left( \omega\right)-\beta_{1}\omega \right) \tilde{E}+i\dfrac{\omega^{2}}{2c^{2}\epsilon_{0}\beta\left( \omega\right)}\tilde{P}_{\mathrm{NL}}, \label{x1}
\end{equation}
where $ z $ is the longitudinal propagation distance, $ \omega $ is the angular frequency,  $ \tilde{E}\left(z,\omega \right)  $ is the spectral electric field, $ \beta_{1}=1/v_{g} $ is the first-order dispersion coefficient, $ v_{g} $ is the group velocity, $ c $ is the vacuum speed of light, $ \epsilon_{0} $ is the vacuum permittivity, $ \beta\left(\omega \right)=\omega/c\left[n_{g} -d/\omega^{2}\right]   $ is the combined gas and fiber dispersion of the fundamental mode (HE$ _{11} $), $ n_{g} $ is the gas refractive index, $ d=\left(2.4048\, c\right)^{2}/2\,r^{2} $, $ r $ is the effective-core radius of the fiber, $\tilde{P}_{\mathrm{NL}} =\mathcal{F}\left\lbrace P_{NL}\left(z,t \right) \right\rbrace $ is the total spectral nonlinear polarization, $\mathcal{F}$ is the Fourier transform, $  t$ is time in a reference frame moving with $ v_{g} $, $ P_{\mathrm{NL}} = P_{\mathrm{K}}+P_{\mathrm{I}}+P_{\mathrm{R}} $, $ P_{\mathrm{K}} $,  $ P_{\mathrm{I}}$, and  $P_{\mathrm{R}} $ are the Kerr, ionization-induced, and Raman nonlinear polarization, respectively, $ P_{\mathrm{K}} =\epsilon_{0} \chi^{\left( 3\right)} E^{3} $, and $ \chi^{\left( 3\right) } $ is the third-order nonlinear susceptibility. We will describe below $ P_{\mathrm{I}} $ and $ P_{\mathrm{R}} $ in more detail.

The time-dependent photoionization-induced polarization satisfies the following evolution equation \cite{Geissler99} 
\begin{equation}
\dfrac{\partial P_{\mathrm{I}}}{\partial t}=\dfrac{\partial n_{e}}{\partial t}\dfrac{U_{I}}{E\left(z,t\right)}+\dfrac{e^{2}}{m_{e}}\displaystyle\int_{-\infty}^{t}\!\!\!\!\!n_{e}\left( z,t'\right)E\left(z,t'\right)dt', \label{x2}
\end{equation}
where $ e $,  $ m_{e} $  are the electron charge and mass, $ U_{I} $ is the ionization energy of the gas under consideration, $ n_{e} $ is the generated free electron density governed by the rate equation
\begin{equation}
\partial_{t} n_{e}=\mathcal{W}\left(z, t\right)\left(n_{\mathrm{T}}-n_{e} \right), \label{x3}
 \end{equation}
 with $n_{ \mathrm{T} }$  the total density of the gas atoms, and  $  \mathcal{W}$ is the ionization rate. Using the Ammosov-Delone-Krainov model, 
\begin{equation}
\mathcal{W}(z,t)=\dfrac{c_{1}}{\left|E\right|^{2n-1}}\, \exp\left[  -c_{2}/\left|E\right|\right]. \label{x4}
\end{equation}
where $ c_{1} =\omega_{p}\zeta\left( 4\omega_{p}\sqrt{2m_{e}U_{I}}/e\right)^{2n-1} $, $ \omega_{p}=U_{I}/\hbar $ is the transition frequency, $\hbar $ is the reduced Planck constant, $ \zeta=2^{2n} /\left[n\Gamma\left( n\right) \Gamma\left( n+1\right) \right] $, $ n=\sqrt{U_{H}/U_{I}} $ is the effective principal quantum number, $ \Gamma $ is the gamma function, and  $ U_{H} =13.6$ eV is the ionization energy of the hydrogen atom.

The dynamics of a specific Raman transition $ l $ in gases can be determined by solving the Bloch equations for an effective two-level system \cite{Kalosha00},
\begin{equation}
\begin{array}{l}
\dfrac{dw}{dt} + \dfrac{w+1}{T_{1}} =\dfrac{i\alpha_{12}}{\hbar}\left(\rho_{12}-\rho_{12}^{*} \right)E^{2},\\
\left[ \dfrac{d}{dt}  + \dfrac{1}{T_{2}}-i\omega_{\mathrm{R}}\right]\rho_{12} =\dfrac{i}{2\hbar}\left[\alpha_{12} w + \left(\alpha_{11}-\alpha_{22} \right)\rho_{12} \right]E^{2},
\end{array} \label{x5} 
\end{equation}
where $ \alpha_{ij} $ and $ \rho_{ij} $ are the elements of the $ 2\times 2 $ polarizability and density matrices, respectively, $ w=\rho_{22}-\rho_{11} $ is the population inversion between the excited and ground states, $ \rho_{22}+\rho_{11}\simeq1 $ is the trace of the density matrix (which states the conservation of probability, assuming long dephasing times, as is the case for many Raman gases),  $ \rho_{21}=\rho_{12}^{*} $, $ \alpha_{12}=\alpha_{21}$,   $ T_{1} $ and $ T_{2} $ are the population and polarization relaxation times, respectively, and the Raman polarization is given by  $ P_{\mathrm{R}}^{\left(l \right) }= \left[\alpha_{11}\rho_{11}+\alpha_{22}\rho_{22}+2 \alpha_{12} \mathrm{Re} \left(\rho_{12}\right)\right] n_{\mathrm{T}} E$. The Raman `coherence' field is defined as $\mathrm{Re} \left(\rho_{12}\right)$. For weak Raman excitation, $ \rho_{11}\approx 1 $ and $ \rho_{22}\approx 0 $, i.e. the second term in $ P_{\mathrm{R}} $ can be neglected, while the first term increases the linear refractive index of the medium by a fixed amount. The total polarization induced by all the excitable Raman modes is  $ P_{\mathrm{R}} = \sum_{l}  P_{\mathrm{R}}^{\left(l \right) }$.

The interplay between the ionization and Raman effects can be investigated during the pulse evolution inside the fiber by solving the strongly nonlinear coupled Eqs. (\ref{x1}-\ref{x5}). We are interested in exploring this interplay when pumping in two different anomalous dispersion regimes: (i) near the ZDW; (ii) away from the ZDW.

\textit{Pumping near the ZDW}, the soliton fission process is expected in this regime due to non-negligible values of high-order dispersion coefficients. Due to the self-phase modulation effect, Raman-induced nonlinear redshift and dispersive wave generation, the spectrum is broadened dramatically and a supercontinuum generation is obtained \cite{Dudley06}. Panels of Fig. \ref{Fig2} show the evolution of an ultrashort Gaussian pulse with full-width-half-maximum (FWHM) 25 fs and centered at 805 nm in a hydrogen-filled HC-PCF with a ZDW at 487 nm.  Hydrogen have two Raman transition modes -- a rotational mode with period 57 fs,  and a vibrational mode with 8 fs \cite{Belli15}. 

The temporal and spectral evolution in the absence of the ionization effects are depicted in Fig. \ref{Fig2}(a,b). Emitted solitons from the soliton-fission process will impulsively excite the rotational Raman mode, since their durations will be shorter than 57 fs.  Hence, they will experience redshift and deceleration in the spectral and time domain, respectively, where the soliton with the shortest duration shows the strongest deceleration and redshift. Also, these solitons will excite a lagging sinusoidal temporal modulation of the medium refractive index that can be thought as a `temporal periodic crystal' \cite{Saleh15a}. The vibrational mode of hydrogen will result in an additional Kerr nonlinearity, since it has an oscillation period less than the duration of the solitons. In Fig. \ref{Fig2}(c), the total Raman coherence field is shown, and is a superposition of two sine waves of different oscillation frequencies, corresponding to the rotational and vibrational Raman frequencies.

The dispersive wave is periodically `twisted' and is radiated during propagation because of the strong non-instantaneous Raman interaction with its emitting soliton. The long relaxation time of Raman in gases (for instance, the rotational mode in hydrogen has $T_{2}=$ 0.43 ns at gas pressure 7 bar and room temperature \cite{Bischel86}) allows the dispersive wave to feel the induced temporal crystal. In analogy with condensed-matter physics, this wave will behave similarly to a wavefunction of an electron in a solid crystal in the presence of a uniform electric field \cite{Saleh15a}. The applied constant force on the temporal crystal arises from the soliton redshift that is accompanied with a uniform acceleration in the time domain. Hence, the twist and the radiation of the dispersive wave can be interpreted as a result of combined Bloch oscillations \cite{Bloch28} and Zener tunneling \cite{Zener34}. In the deep-UV, there is an emission of a very-weak dispersive wave (faint line on the right of the spectrum) due to interaction between positive and negative frequency components of the electric field \cite{Conforti13}.

Switching on the ionization nonlinearity does not change the qualitative picture of the spectrum evolution, as shown in panel (d). However, the spectral broadening  is reduced due to the ionization loss that suppresses the pulse intensity due to plasma formation \cite{Hoelzer11b,Saleh11a}. Due to photoionization effects, the very-weak dispersive-wave in the deep UV disappears, and another wave in the mid-infrared (faint line on the left of the spectrum) is emitted instead, due to the photoionization-induced dispersion \cite{Novoa15}.

\begin{figure}
\includegraphics[width=8.6cm]{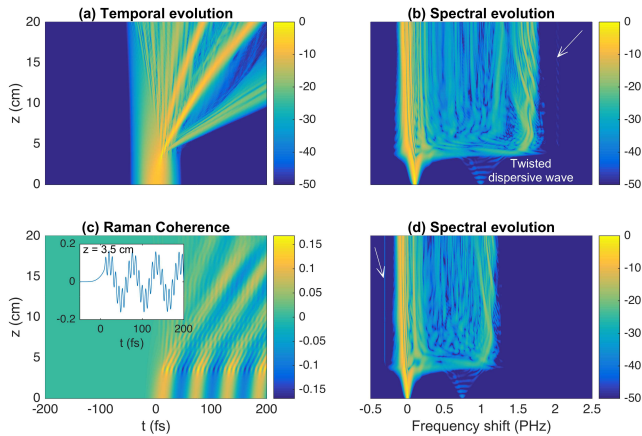}  
\caption{(Color online). (a) Temporal and (b) spectral evolution of a Gaussian pulse propagating with central wavelength 0.805 $ \mu $m, energy 3  $ \mu $J and FWHM 25 fs in a H$ _{2} $-filled HC-PCF with a gas pressure 7 bar, and a hexagonal core-diameter 27 $ \mu $m in absence of ionization effects. (c) Temporal evolution of Raman coherence wave induced by the rotational and vibrational excitations. The inset shows the total Raman coherence at $z = 3.5$ cm. (d) The effect of ionization on the spectral evolution of the Gaussian pulse, using the ionziation energy of the hydrogen molecule $U_{I} =15.4$ eV \cite{Saenz02}. Contour plots in (a), (b), and (d) are given in a logarithmic scale and truncated at -50 dB. White arrows point to the weak (b) UV negative, and (c) IR dispersive waves. The linear, nonlinear, and Raman parameters of hydrogen can be found in \cite{Belli15}. \label{Fig1}}
\end{figure}

\begin{figure}
\includegraphics[width=8.6cm]{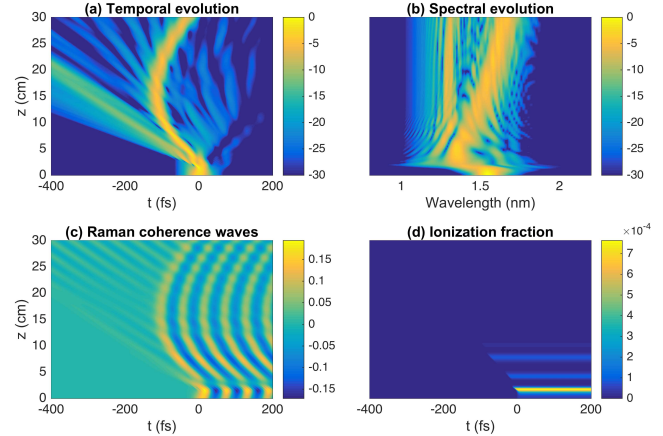}
\caption{(Color online). (a) Temporal and (b) spectral evolution of a sech-pulse with central wavelength 1.55 $ \mu $m, energy 2.7  $ \mu $J and FWHM 30 fs in a H$ _{2} $-filled HC-PCF with a gas pressure 7 bar, and an effectivel core-diameter 18 $ \mu $m. Temporal evolution of the (c) Raman coherence waves induced by rotational and vibrational excitations, and (d)  ionization fraction. Contour plots in (a) and (b) are given in a logarithmic scale and truncated at -30 dB. \label{Fig3}}
\end{figure}

\begin{figure}
\includegraphics[width=8.6cm]{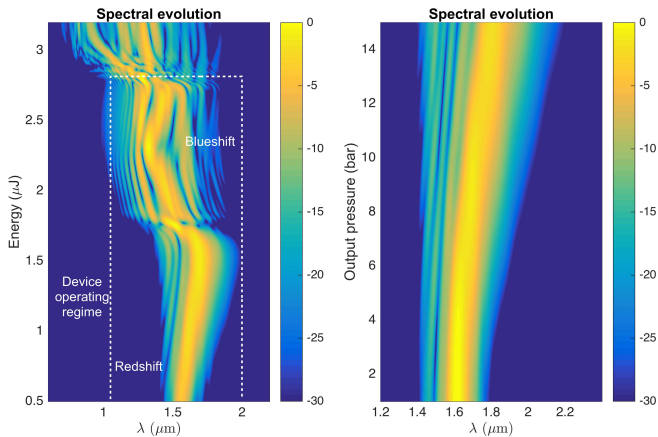}
\caption{(Color online). (a) Energy and (b) output-pressure dependencies of the final spectral profiles of different sech-pulses with central wavelength 1.55 $ \mu $m, and FWHM 30 fs in H$ _{2} $-filled HC-PCFs with length 10 cm,  input gas pressure 7 bar, and an effectivel core-diameter 18 $ \mu $m.  (a) The input pulse energy is scanned from 0.5 to 3.2 $ \mu $J, and the output pressure is fixed at 7 bar.  (b) The output pressure is scanned from 1 to 15 bar, and the input pulse energy is fixed at 1.3 $ \mu $J,.  Contour plots are given in a logarithmic scale and truncated at -30 dB. \label{Fig2}}
\end{figure}

\textit{Working in the deep-anomalous regime}, i.e. far from the ZDW, allows the observation of a soliton that can have a clean self-frequency shift, without emitting a dispersive wave that lies far-way from the soliton central wavelength. Hence, we expect that in this regime the interplay between photoionization and Raman effects can lead to interesting features. The evolution of a sech-pulse with an input energy  2.7 $ \mu $J and centered at  1.55 $ \mu $m in a 25 cm long H$ _{2} $-filled PCF and a ZDW at 403 nm is depicted in Fig. \ref{Fig3}. Panels (a,b) show the temporal and spectral evolution of the pulse, while panels (c,d) show the temporal evolution of the Raman coherence and the ionization fraction. The net self-frequency shift depends on whether photoionization or Raman nonlinearity is dominant. Photoionizaton-induced blueshift occurs only when pulse intensity exceeds the ionization threshold \cite{Saleh11a}. However, Raman-induced redshift takes place along the whole propagation with a rate that has an approximately linear-dependence on the pulse intensity \cite{Saleh15a}. Initially, the pulse is intense enough to ionize the gas and generates enough electrons that contribute to the pulse blueshift. As the pulse shifts to shorter wavelengths,  the group velocity dispersion decreases and the Kerr nonlinearity increases. This results in an adiabatic pulse compression \cite{Chang13} that increases the amount of ionization  and blueshift. As clear from panel (b), photoionization is dominant during the first propagation stage along the fiber. However due to the concurrent ionization losses, the blueshift process ceases after a certain propagation distance $ z_{\mathrm{Ion}} $, allowing the redshifting process to be dominant in the rest of the fiber. Panels (a,c) depict how the pulse and the Raman-induced coherence switch between acceleration and deceleration. The fiber length is a key-element in determining the net frequency shift at the fiber output.

To explore the strong interaction between Raman and photoionization effects in this regime, the energy-dependency of the output spectral profile of an initial sech-pulse centered at  1.55 $ \mu $m is portrayed in Fig. \ref{Fig2}(a). For small input energies, ionization threshold that allows for plasma generation is not reached. Hence, Raman nonlinearity is dominant; its vibrational mode modifies the Kerr nonlinearity, while its rotational mode introduces linear self-frequency redshift. Around 1.75 $ \mu $J photoionization-induced self-frequency blueshift starts to take place due to plasma formation. As the input pulse energy increases, the induced blueshift starts to compensate then take over the rotational Raman-induced redshift. Initially, the blueshift increases linearly with the input energy, then there is a slight decrease beyond about 2.4  $ \mu $J. This is because the photoionization process at this latter range of energies occurs very close to the fiber input, allows the Raman process to almost be dominant along the whole fiber, resulting in enhancing the reshift of the spectrum. Increasing the input energy more and more, the pulse breaks up and the clear soliton self-frequency shift is suppressed. The unambiguous interplay between photoionization and Raman processes for small input energies in this case  suggests the proposal of a novel tunable device that can be used for either frequency-down or frequency-up conversion. The pulse central frequency can be scanned over 400 nm range from about 1.3 -- 1.7 $ \mu $m by changing its input energy over the range 0.5 -- 2.8 $ \mu $J.

The ratio between the fiber length and $ z_{\mathrm{Ion}} $ is a key-element in  determining the frequency-up/down conversion range. Increasing the fiber length will certainly increase the frequency-down conversion, since the Raman nonlinearity will become dominant eventually. The latter is because the ionization loss halts the blueshift after the first few centimeters of the fiber and allows the redshift process to rule in the remaining distance. Contrarily, decreasing the fiber length, the frequency-up conversion range is enhanced over the frequency-down conversion range. So, there is a trade-off for choosing the length of the proposed device. However, the flexibility of tuning the output pressure at the gas-filled fiber end offers an additional degree of freedom to overcome this problem \citep{Suda05}. At equilibrium, the pressure distribution across the fiber is $ P\left(z \right)=\left[ P_{0}^{2}+\frac{z}{L} \left( P_{L}^{2}-P_{0}^{2}\right)\right] ^{1/2}  $, where $ P_{0} $ and $ P_{L} $ are the input and output pressures, and $ L $ is the fiber length. Panel (b) of Fig. \ref{Fig2} shows the dependence of the final spectrum of an initial sech-pulse on the output pressure with a fixed input pressure and pulse energy. By increasing the output pressure, the spectrum is linearly shifted towards the red side. For instance, by tuning the output pressure from 1 -- 15 bar, the spectrum is redshifted by about 250 nm. This allows to shorten the fiber length  close to $ z_{\mathrm{Ion}} $  to enhance the photoionization-induced blueshift. Then, the reduced redshift at lower input energies can be recouped by raising the output-fiber pressure. This  enables the proposed device to have a relatively wide tunable range for both frequency-up/down conversions.
 
In conclusion, we have investigated the interplay between the nonlinear photoionization and Raman effects in gas-filled HC-PCFs in two different dispersion regimes, pumping near or away from the ZDW. In the former regime, photoionization tends to degrade the  broadening of the supercontinuum generation due to the ionization loss. However in the latter regime, photoionization induces a clear self-frequency blueshift that works against the Raman-induced redshift. This strong interplay suggests the proposal of a novel device that can be used for either frequency-up/down conversion via tuning the input pulse energy and the gas-pressure at the fiber output.

The authors would like to thank Dr. Wonkeun Chang at the Australian National University as well as researchers at Russell Division in Max Planck Institute for the Science of Light in Germany  for useful discussions. M. Saleh would like also to acknowledge deeply the support of this research by the Royal Society of Edinburgh and Scottish Government.

\bibliographystyle{apsrev4-1}	

\end{document}